\documentclass[a4paper,11pt]{article}
\pdfoutput=1 

\usepackage{jinstpub} 

\usepackage{subfigure}
\usepackage{floatrow}
\newfloatcommand{capbtabbox}{table}[][\FBwidth]
\usepackage{blindtext}
\usepackage{adjustbox}
\usepackage{floatrow}
\usepackage{lineno} 

\title{\boldmath Performance of the ALICE Time-Of-Flight detector at the LHC}

\author[a,b,c,1] {F.~Carnesecchi \note{Corresponding author.}}
 \affiliation[a] {Museo Storico della Fisica e Centro Studi e Ricerche "Enrico Fermi'', \\Roma, Italy}
 \affiliation[b]{Dipartimento di Fisica e Astronomia dell'Universit\`a, \\Bologna, Italy}
 \affiliation[c]{Istituto Nazionale di Fisica Nucleare, \\ Sezione di Bologna, Italy }

\emailAdd{francesca.carnesecchi@bo.infn.it}
\collaboration[c]{on behalf of the ALICE Collaboration}

\abstract{
The ALICE Time-Of-Flight (TOF) detector at LHC is based on the Multigap Resistive Plate Chambers (MRPCs). The TOF performance during LHC Run 2 is here reported. Particular attention is given to the improved time resolution reached by TOF detector of $56$ ps, with the consequently improved particle identification capabilities.}


\arxivnumber{} 

\begin{document}

\maketitle
\flushbottom

\section{Introduction}
\label{sec:intro}
The ALICE \cite{ALICE2008,ALICE2014} experiment at CERN has been designed to study the properties of the strongly interacting, dense and hot matter created in ultra-relativistic Pb--Pb collisions. 
Two running periods have been performed, Run 1 (2009--2013) and Run 2 (2015--present), with different centre-of-mass energies and collision systems  (pp, p--Pb, Pb--Pb).
Particle IDentification (PID) is essential, as many observables are either mass or flavour dependent.
ALICE performs PID making use of almost all known techniques, in a range of momenta between $0.15~\text{GeV}/c$ to 20 GeV/$c$ \cite{ALICE2014,Cifarelli_2016}.\\
We will focus on the Time-Of-Flight (TOF) detector of ALICE \cite{TOF_tech,TOF_techAdd} that is located at about $3.7~\text{m}$ from the beam axis and covers with a large cylindrical array ($\sim141 ~\text{m}^2$ of active area) the central region ($-0.9 < \eta< 0.9$).
The TOF provides charged-particle PID in the intermediate momentum range. 
The TOF also provides a trigger for cosmic ray events and ultraperipheral collisions.
The Multigap Resistive Plate Chamber (MRPC) strip detector is the basic unit of the ALICE TOF detector, with a $120 \times 7.4~ \text{cm}^2$ active area.
The ALICE TOF array consists of $1593$ MRPC strips, subdivided into 18 azimuthal sectors. 
To guarantee low detector occupancy even in the highest charged-particle-density scenario each MRPC strip is segmented into two rows of 48 pickup pads of $3.5 \times 2.5 ~\text{cm}^2$, for a total of 96 pads for each strip and 152928 total readout channels.
The TOF MRPC is based on a double-stack design: it is made of two stacks of five gas gaps. The resistive plates are made with commercially available soda-lime glass sheets. The gap (250 $\mu$m) is realised by commercial fishing line stretched across the glass sheets. 
In a beam test setup \cite{tofBeam2004} the average MRPC time resolution, including the contributions of the complete front-end and readout electronics, was measured to be better than 50 ps.\\
In this paper the TOF performance during the Run 2 data taking period at the LHC is discussed in detail, with several comparisons to the TOF Run 1 performance, reported in \cite{TOFperform2013}.
\section{Operation}
The TOF, despite being from almost ten years in operation (installed in 2008, taking data from the first LHC collisions), is showing no signs of degradation. In 2017 the TOF was operated for a total of 2116 hours, corresponding to a $99\%$ of the total time availability (with the Time Projection Chamber running). The average fraction of active channels was $93\%$, with the remaining $7\%$ being not available due to faults in the readout electronics ($4\%$) or in the external HV distribution ($3\%$).  
 \begin{figure}[htbp]
        \centering%
        \subfigure[\label{fig:TOFcurrentNoBeam}]%
          {\includegraphics [width=8.5cm] {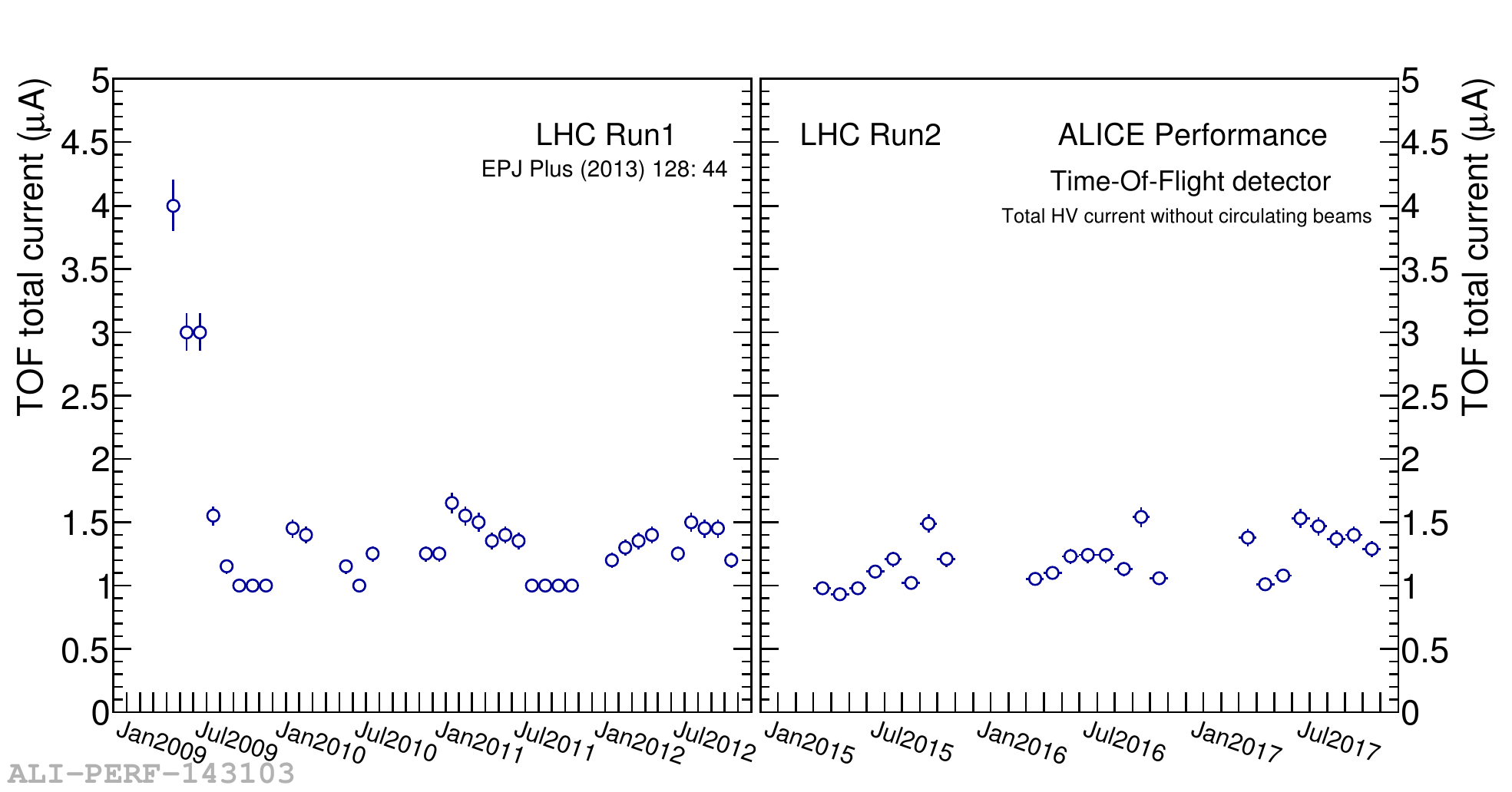}}\quad
        \centering%
        \subfigure[\label{fig:TOFcurrentVsLumi_logs}]%
          {\includegraphics[width=9cm]{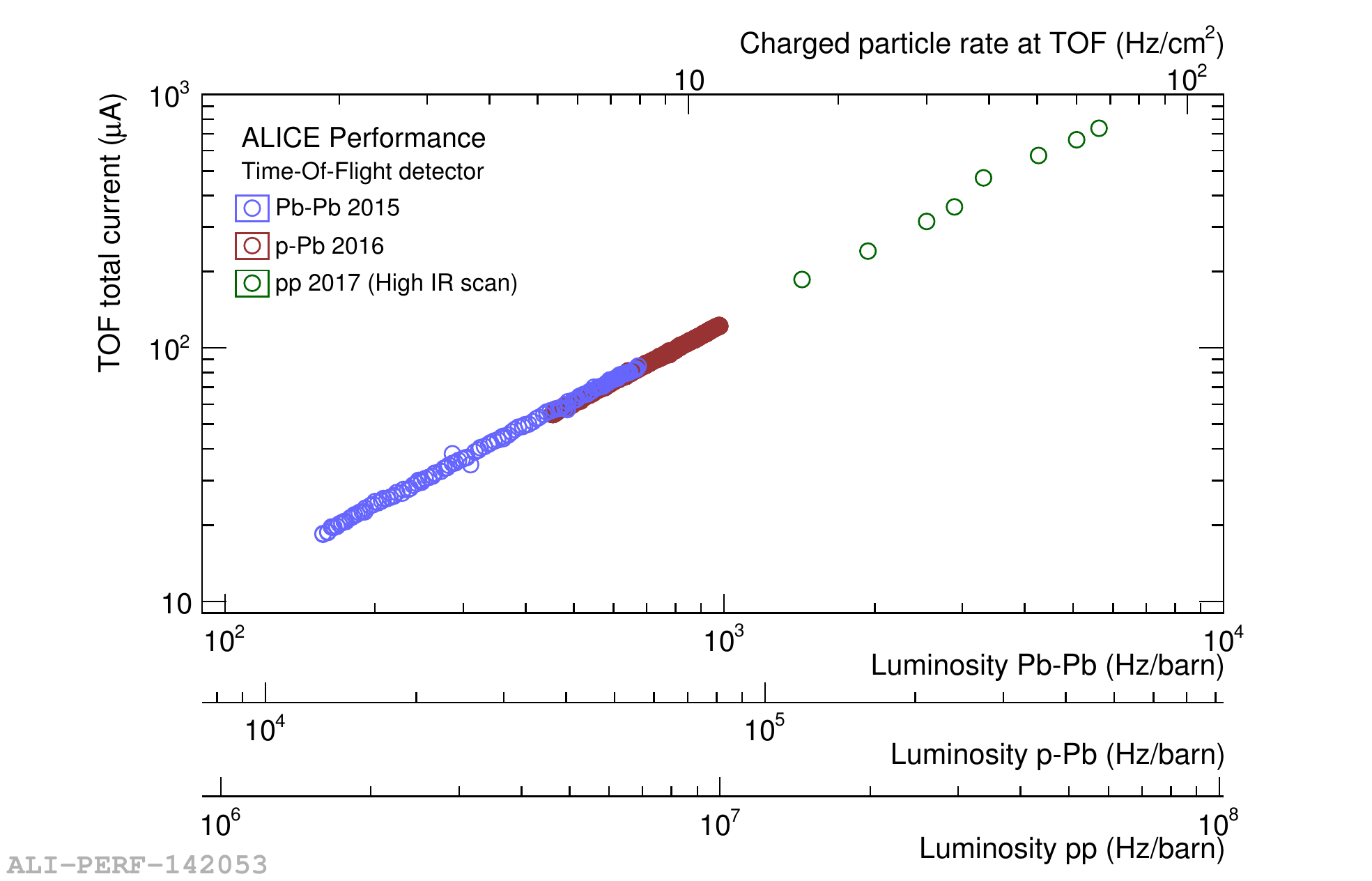}}
        \caption{(a) TOF total current measured without circulating beam as a function of time. The initial drop is due to an initial MRPCs HV conditioning at the beginning of the installation. (b) TOF total current as a function of the measured luminosity on the bottom x-axis and the calculated rate at TOF on the top x-axis. The luminosity for the different collision systems are aligned to the same detector load. The points at high rate are obtained with a dedicated scan. The average charge for a track is about $6~\text{pC}$, compatible with  \cite{TOFperform2013}. 
.\label{fig:TOFcurrent}}
\end{figure}

\subsection{MRPC total current: with or without beam}
The voltage on each MRPC is applied symmetrically $\pm 6500$ V, w.r.t. the middle of the stacks. The TOF total current is defined as the sum of the current over all the 1593 MRPCs. In Fig.~\ref{fig:TOFcurrentNoBeam} the total current, without circulating beam, versus time is reported. 
A total current of about $1~\mu \text{A}$ has been measured, namely about $1~\text{nA}$ for each MRPC; such a low current is also reflected in the low detector noise. The low TOF current and its stability do not show signs of degradation or ageing of the detector over the years of operations.
In Fig.\ref{fig:TOFcurrentVsLumi_logs} the total current versus the instantaneous luminosity at LHC is reported. 
The current increases linearly with the luminosity, making the TOF a luminometer detector. There are no signs of deviations, even at the high luminosities expected for Run 3 Pb--Pb collisions of 5000 Hz/barn. The corresponding Run 3 rate at the TOF surface of  $60~\text{Hz/cm}^2$ is well below the limit of $1~\text{kHz/cm}^2$ where the MRPC performance was observed to decrease \cite{Akindinov2002}. 
The maximum total current reached was $\sim 700~\mu \text{A}$, namely $0.4~ \mu \text{A}$ for each MRPC.

\subsection{Matching efficiency}

\begin{figure}[htbp]
        \centering%
        \subfigure[\label{fig:TOFME_vsPt-2013-2016}]%
          {\includegraphics [height=4.5cm, width=6cm] {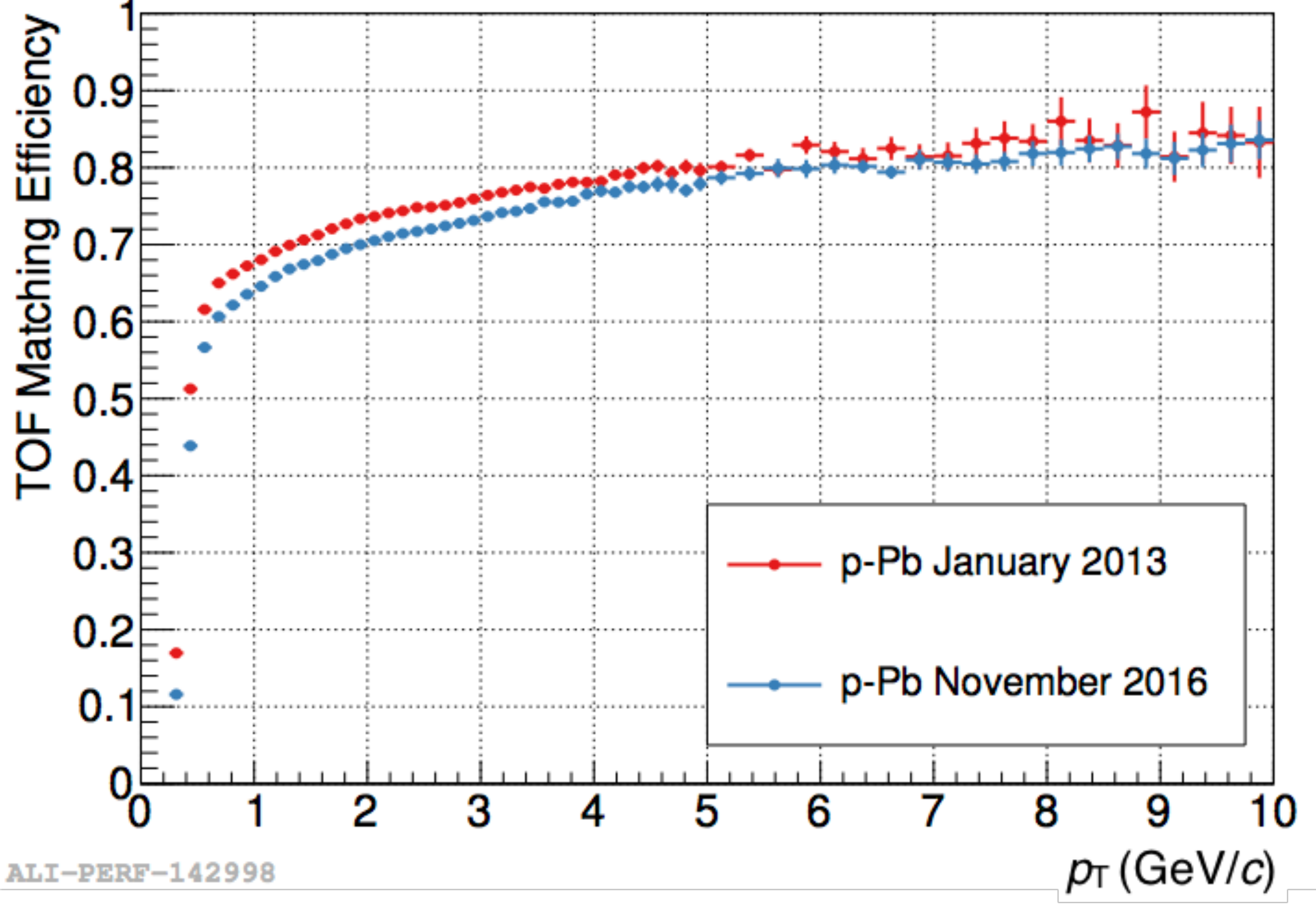}}\quad
        \centering%
        \subfigure[\label{fig:TORCosmicTriggerRate}]%
          {\raisebox{0.09cm}{\includegraphics[height=4.8cm, width=6.2cm]{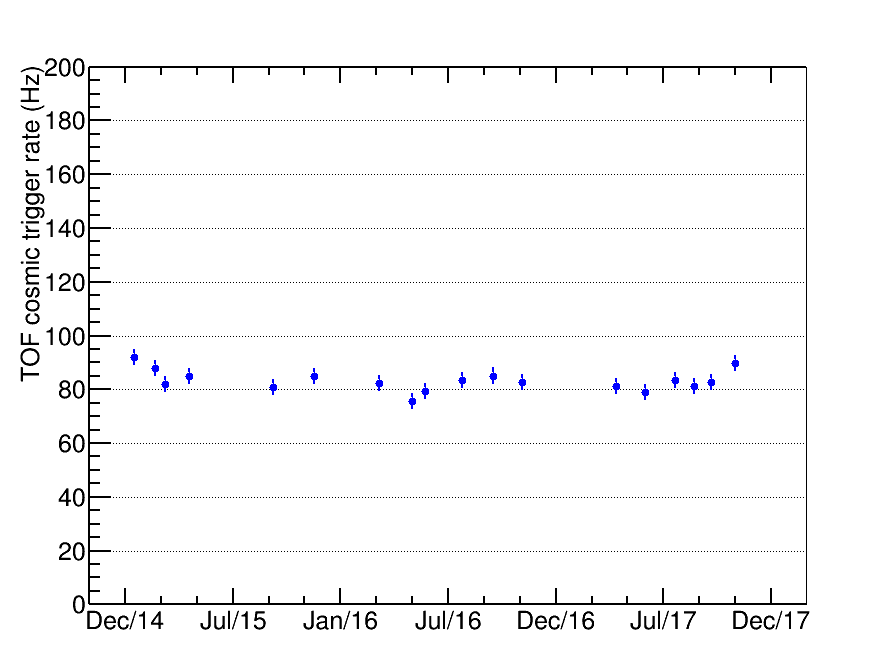}}}
        \caption{(a) TOF matching efficiency versus the transverse momentum for p--Pb collision in 2013 (Run 1) and 2016 (Run 2). (b) TOF cosmic ray trigger rate versus time (Run 2).\label{fig:run1-2}}
\end{figure}

In Fig.\ref{fig:TOFME_vsPt-2013-2016} it is reported the TOF matching efficiency as a function of the transverse momentum, compared between 2013 (Run 1) and 2016 (Run 2). The matching efficiency is defined as the number of tracks matched with a signal on the TOF divided by the number of tracks reconstructed in the inner detectors. The values of the efficiency depend from several factors: 
the intrinsic MRPC detector efficiency ($\sim98-99\%$~\cite{TOFperform2013}),
the efficiency of the TOF reconstruction algorithm,
the TOF geometrical acceptance (dead space), 
the material budget in front of the TOF,
and the accuracy in the track extrapolation from the outer layers of the tracking detectors to the TOF surface.
At low momenta the matching efficiency goes to zero, as tracks with transverse momentum lower than 0.3 GeV/$c$ are bent in the ALICE magnetic field and do not reach the TOF. 
Since between Run 1 and Run 2 the material budget in front of the TOF increased (with the installation of the missing TRD modules) and the algorithm of charged-track reconstruction was modified in the tracking detectors, a lower efficiency is observed at lower momenta (below $4~$GeV/$c$) in the 2016 sample. 

\subsection{TOF cosmic and collision triggers}
Thanks to its fast signal, low noise, high granularity and large coverage of the detector, TOF is used as trigger detector both for cosmic-ray events and for resonances produced in Ultra-Peripheral Collisions (UPC). For the UPC, the expected topology consists of two tracks in the central barrel detectors with forward detectors showing no activity.
For the cosmic rays, the topology of the trigger is back to back; a trigger is sent when there is a coincidence between two opposite sectors in the azimuthal plane. 
In Fig.\ref{fig:TORCosmicTriggerRate} the cosmic ray rate from 2014 until 2017 is reported which shows a remarkable stability over all the period, with an average of about $85~\text{Hz}$.\\

\section{Time performance and calibrations}
The gaussian distribution used for the PID ($t_{\text{TOF}}-t_{\text{event}}-t_{\text{trk}_\text{i}}$, see Sec.\ref{sec:PID}, with $t_{\text{event}}$ the event time of the collision and $t_{\text{trk}_\text{i}}$ the expected arrival time for different mass hypotheses i= $\pi, \text{K}, \text{p},\dots$) should be centred at zero with a $\sigma_{\text{TOT}}$ standard deviation:
\begin{align}
\label{eq:res_TOF}
\sigma^2_{\text{TOT}}&= \sigma^2_{\text{TOF}}+\sigma^2_{\text{trk}}+\sigma^2_{\text{event}} \\
&= \sigma^2_{\text{MRPC}}+2~ \sigma^2_{\text{TDC}}+\sigma^2_{\text{FEE}}+\sigma^2_{\text{clock}}+\sigma^2_{\text{Cal}}+\sigma^2_{\text{trk}}+\sigma^2_{\text{event}}
\end{align}
with $\sigma_{\text{trk}}$ the tracking particle jitter\footnote{After the particle is reconstructed by other ALICE detectors, it must be extrapolated to the active area of TOF;  here a matching window of $3 ~\text{cm}$ for Pb--Pb collisions ($10$ cm for pp collisions) is opened around the extrapolation point. The matching algorithm then searches for at least one signal and the closest one is combined to the track. 
}, $\sigma_{\text{event}}$ the resolution on the event time (see Sec.\ref{subsec:eventtTme}),
$\sigma_{\text{MRPC}}$ the MRPC intrinsic resolution, $\sigma_{\text{TDC}} (\sim$$20$ ps) and $\sigma_{\text{FEE}} (\sim$$10$ ps) the readout and front-end electronic time resolution, $\sigma_{\text{clock}}(\sim$$15$ ps) the LHC clock, and $\sigma_{\text{Cal}}$ the calibration time resolution. The calibration procedure is based on three contributions: a common offset, that captures the average phase shift between the LHC clock and the actual collision time, a channel-by-channel offset, which compensates for the different time delays introduced by the cable routing from the MRPC to the TDC channels, and a channel-by-channel time-slewing correction. 
 \begin{figure}[htbp]
        \centering%
        \subfigure[\label{fig:res_TOF}]%
          {\includegraphics [height=5.6 cm] {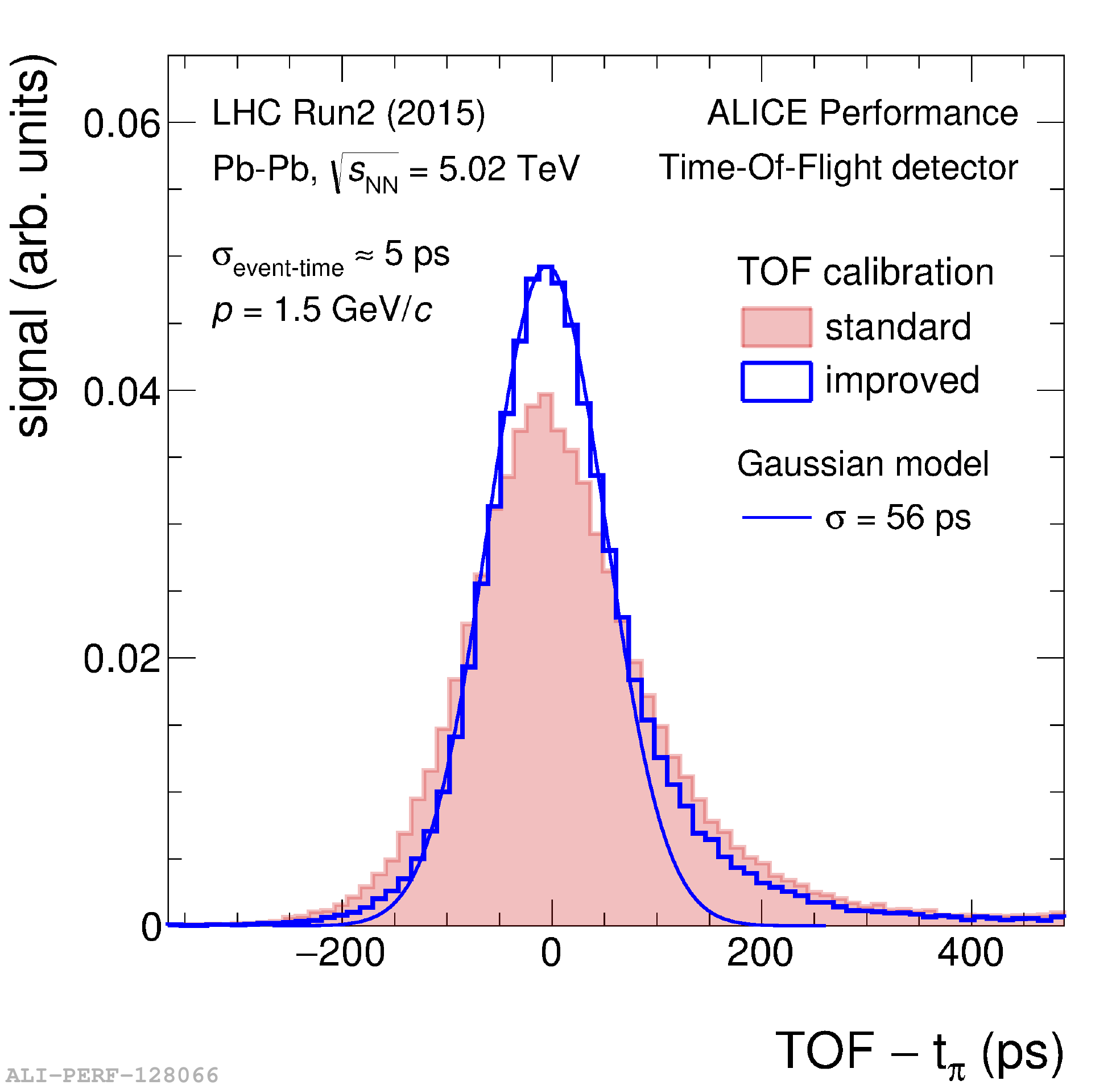}}\quad
        \centering%
        \subfigure[\label{fig:slewing-1_Vsold}]%
          {\raisebox{0.25cm}{\includegraphics[height=5.1 cm]{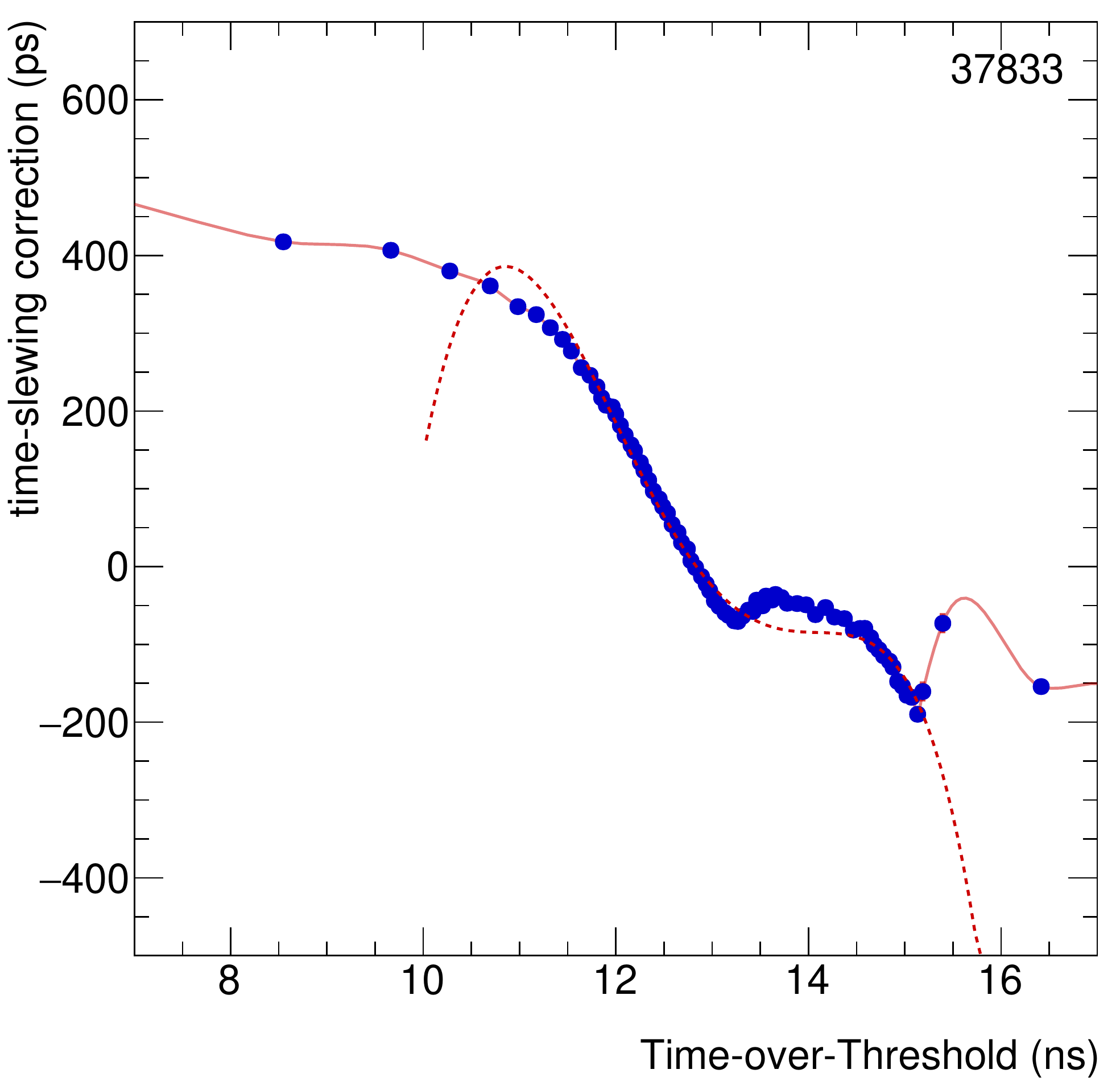}}}
        \caption{(a) TOF time resolution of $56$ ps in Pb--Pb collisions at $5.02$ TeV. 
        Along the x-axis, $\text{TOF}-t_{\pi}$ corresponds to $t_{\text{TOF}}-t_{\text{event}}-t_{\text{trk}_{\pi}}$. (b) Time distribution versus the TOT (Time Over Threshold) for the channel number 37833. The continuous red line represents the improved algorithm while the dotted line correspond to the old 5th-order-polynomial-based calibration. \label{fig:run1-2}}
\end{figure}
There are also other factors that can lead to a worsening of time resolution: the time walk effect (a delay time due to the finite size of the readout pad), the hit multiplicity (when a particle passes near the edge of a readout pad it can induce a signal in more than one pad (cluster) with a consequently worse time resolution) and presence of the asymmetric tail, visible in the time distribution. While the first two factors are being analysed and corrected for, the asymmetric tails are still under study and a complete understanding of their origin is still not settled.
Thanks to the new calibration campaign performed in 2017 (see Sec.\ref{subsec:tscorr}), the ALICE TOF time resolution has been improved reaching the value of about $56$ ps. 
In Fig.\ref{fig:res_TOF} the time distribution is reported compared with the old one (standard) of about $84$ ps \cite{TOFperform2013}. 

\subsection{Time-slewing corrections}
\label{subsec:tscorr}
The time-slewing correction is necessary to disentangle the time from the charge information; indeed, due to the comparison of the signal with a fixed threshold, the time measurement is correlated to the signal amplitude.
In the ALICE TOF system the Time-Over-Threshold (TOT) is used as a proxy for the signal charge amplitude. Presently the time-slewing correction leads to an improvement of $50 \%$ on the TOF time resolution. Thanks to a new calibration campaign which made use of a much larger data sample, a channel-by-channel time-slewing calibration could be performed. This is a significant improvement with respect to the previous time-slewing calibration  \cite{TOFperform2013}, which was based on a grouping of 8 channels using a fitted parameterisation of the time-TOT correlation in a limited TOT range. 
In Fig.\ref{fig:slewing-1_Vsold} the time of arrival versus the TOT for a single channel is reported. The continuous red line represents the improved correction while the dotted line represents the old algorithm. 
Moreover, the calibration is found to be very stable with time (in 2017 the $88\%$ of the channels show a RMS smaller than $20$ ps).

\subsection{Event time determination with TOF}
\label{subsec:eventtTme}
Due to the finite dimensions of the bunches that collide at LHC, $t_{\text{event}}$ is approximately gaussian distributed with width $\sigma_{\text{event}}$$\sim80-200$ ps. The ALICE T0 detector is designed to provide an accurate time measurement of $t_{\text{event}}$. Since the T0 detector has a limited acceptance the information is only available in a fraction of the events. A method based on combinatorial minimisation algorithm to measure the $t_{\text{event}}$ has been developed by using the TOF itself ($t^{\text{TOF}}_{\text{ev}}$) \cite{TOF_T0_2017}. 
The final $t_{\text{event}}$ is a weighted average with time resolution between T0 and TOF algorithms, when both are available.
 \begin{figure}[htbp]
        \centering%
        \subfigure[\label{fig:t_event_Run1_Run2}]%
          {\includegraphics [height=5.cm] {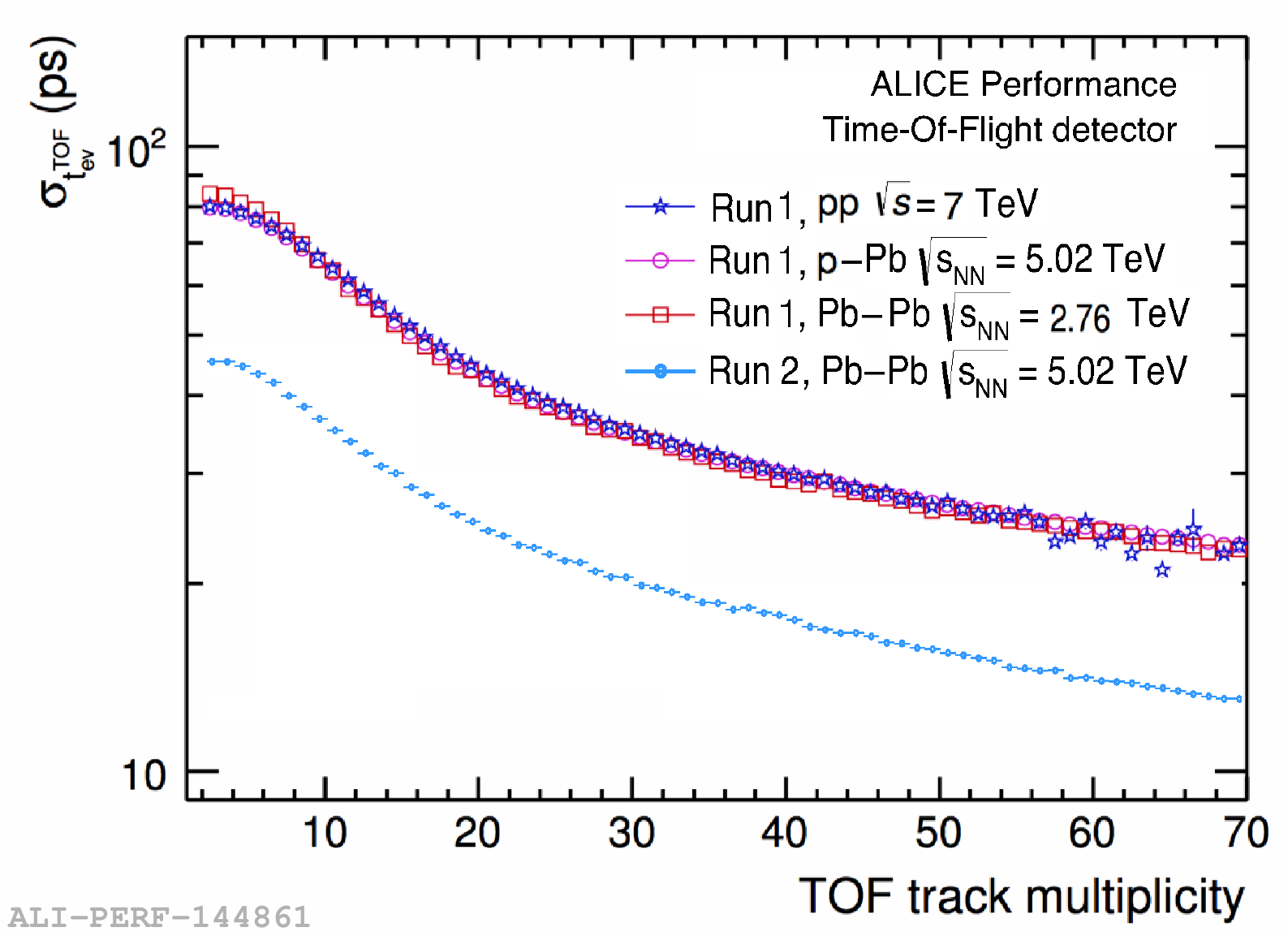}}\quad
        \centering%
        \subfigure[\label{fig:t_ev_TOF}]%
          {\raisebox{0.0cm}{\includegraphics[height=5.14 cm, width=7.cm]{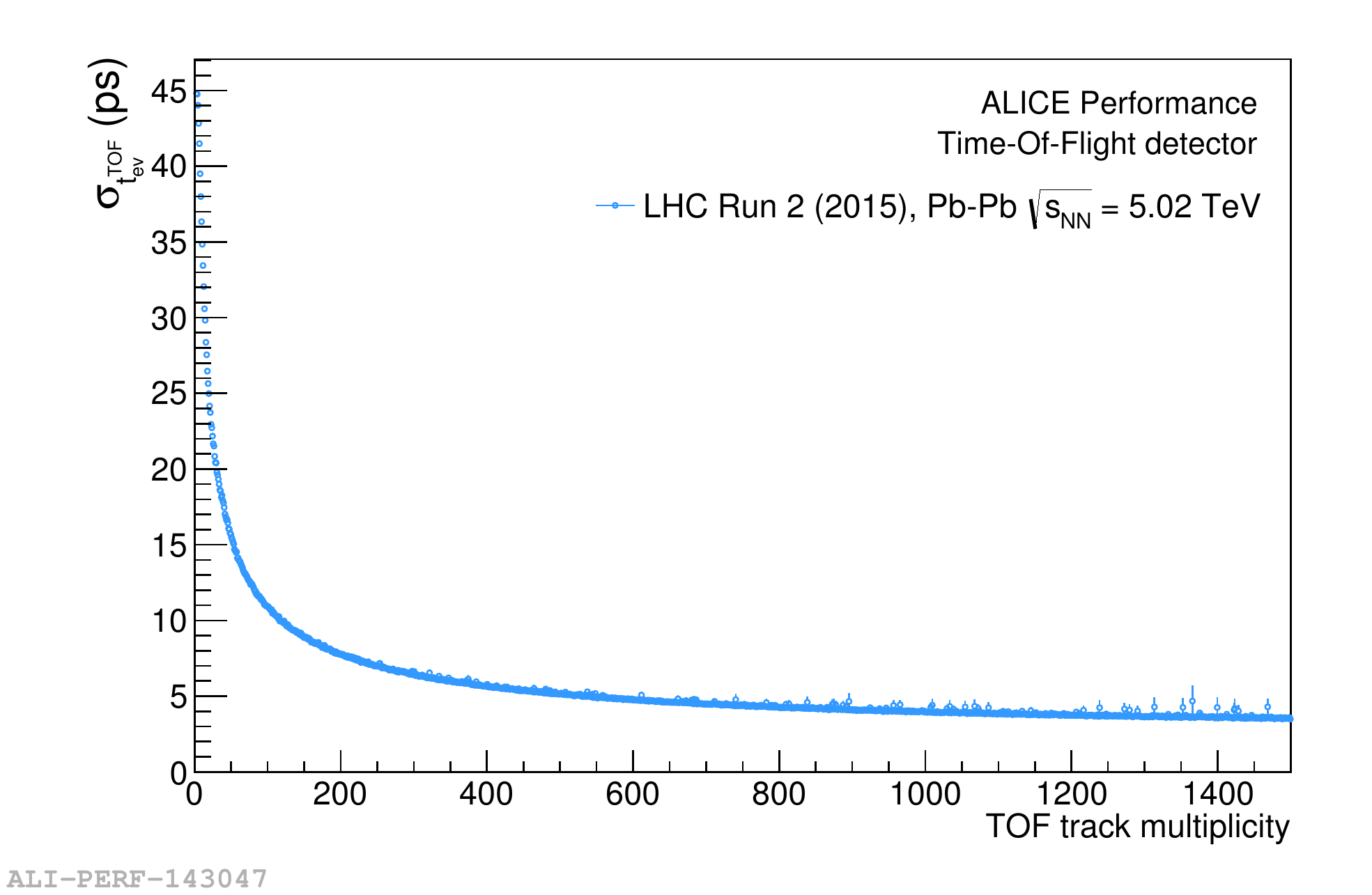}}}
        \caption{(a) Resolution of $t^{\text{TOF}}_{\text{ev}}$ as a function of the TOF track multiplicity for different collision systems (pp, p--Pb, Pb--Pb) and at different periods: Run 1 and Run 2 (or rather before and after the improved calibration). (b) Same data of the Run 2 reported in Fig.\ref{fig:t_event_Run1_Run2} demagnified to cover a larger range of TOF track multiplicity.
\label{fig:run1-2}}
\end{figure}
In Fig.\ref{fig:t_event_Run1_Run2} the time resolution of $t^{\text{TOF}}_{\text{ev}}$ versus the TOF track multiplicity is reported for Run 1 and Run 2. 
The TOF track multiplicity of the event is the number of tracks matched with a hit on the TOF detector\footnote{It should be pointed out that the TOF track multiplicity does not represent the number of tracks that are used by the TOF algorithm to compute the $t^{\text{TOF}}_{\text{ev}}$ ; the number of the used TOF tracks is indeed lower since in the algorithm a further basic selection on the quality of the track is applied to guarantee a good quality of the $t^{\text{TOF}}_{\text{ev}}$ \cite{TOF_T0_2017}.}.
From Run 1 data it can be concluded that the resolution is independent from the collision system (pp, p--Pb, Pb--Pb). From the comparison of Run 1 and Run 2 the improvement due to the new calibrations is visible. It should be noted that $t^{\text{TOF}}_{\text{ev}}$ improves with higher track multiplicity as
\begin{equation}
\label{eq:tevtof}
t^{\text{TOF}}_{\text{ev}} \propto \frac{\sigma_\text{TOT}}{\sqrt{\text{~TOF track multiplicity}}}
\end{equation} 
In Fig.\ref{fig:t_ev_TOF} a demagnified version of the previous plot is reported. For a multiplicity larger than 100 ( $0$--$50\%$ of centrality) the $t^{\text{TOF}}_{\text{ev}}$ is lower than $10$ ps.

\section{Particle IDentification with TOF}
\label{sec:PID}
\begin{figure}[htbp]
        \centering%
        \subfigure[\label{fig:TOFBetawoMismatchEtaCut_PbPb}]%
          {\includegraphics [height=5.14cm] {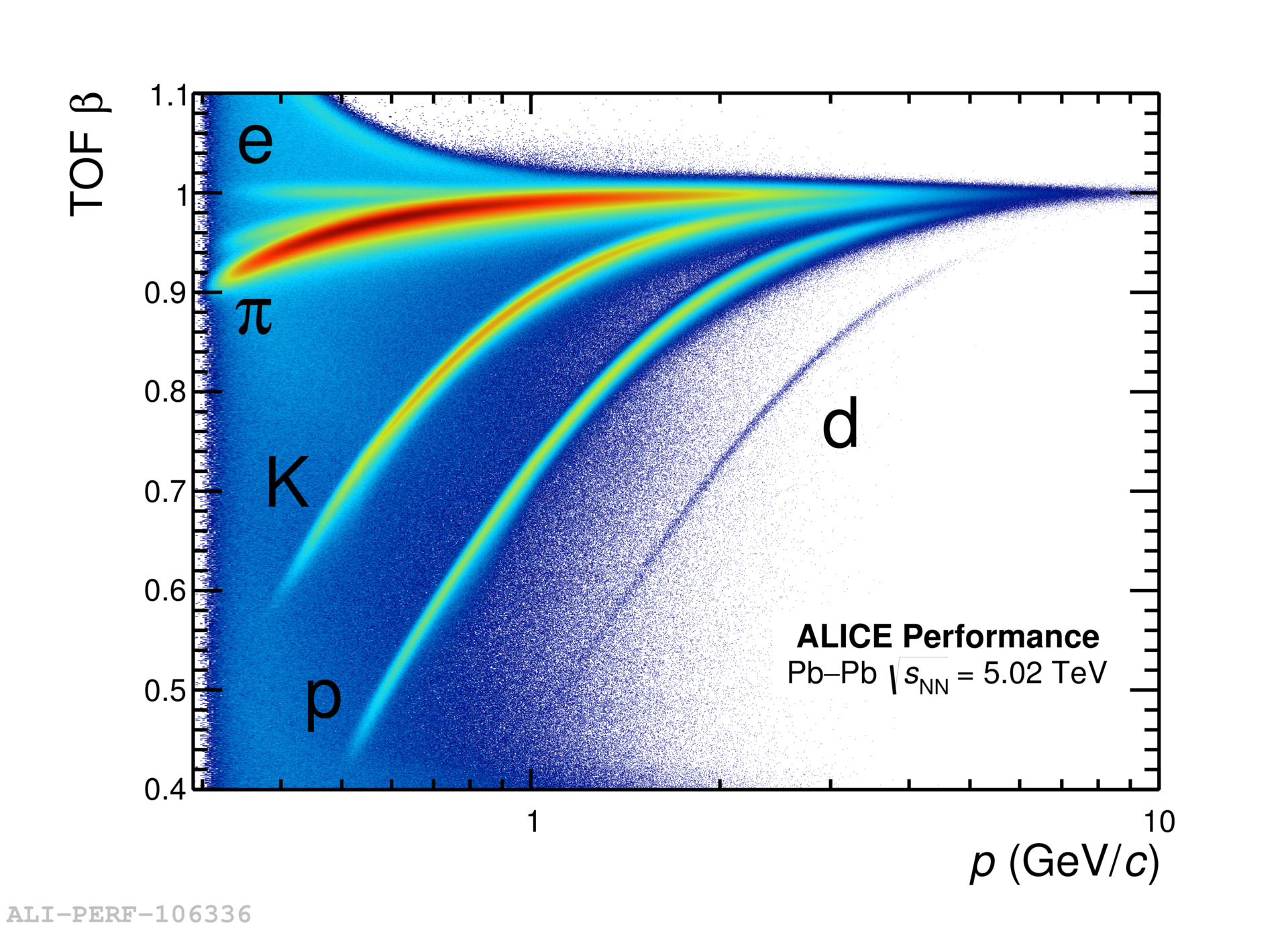}}\quad \quad\quad \quad
        \centering%
        \subfigure[\label{fig:TOFBeta_Cut3_PbPb}]%
          {\raisebox{0.0cm}{\includegraphics[height=5. cm, width=6.9cm]{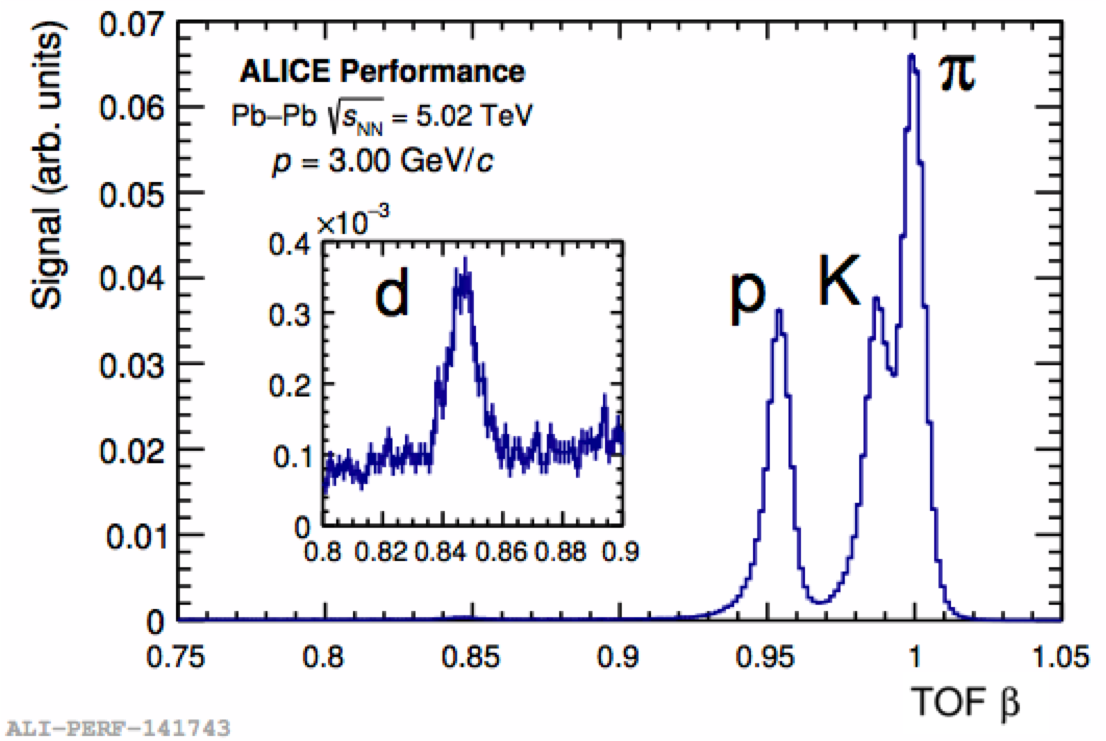}}}
	 \centering%
        \subfigure[\label{fig:TOFBeta_Cut5_PbPb}]%
          {\raisebox{0.0cm}{\includegraphics[height=5. cm, width=6.9cm]{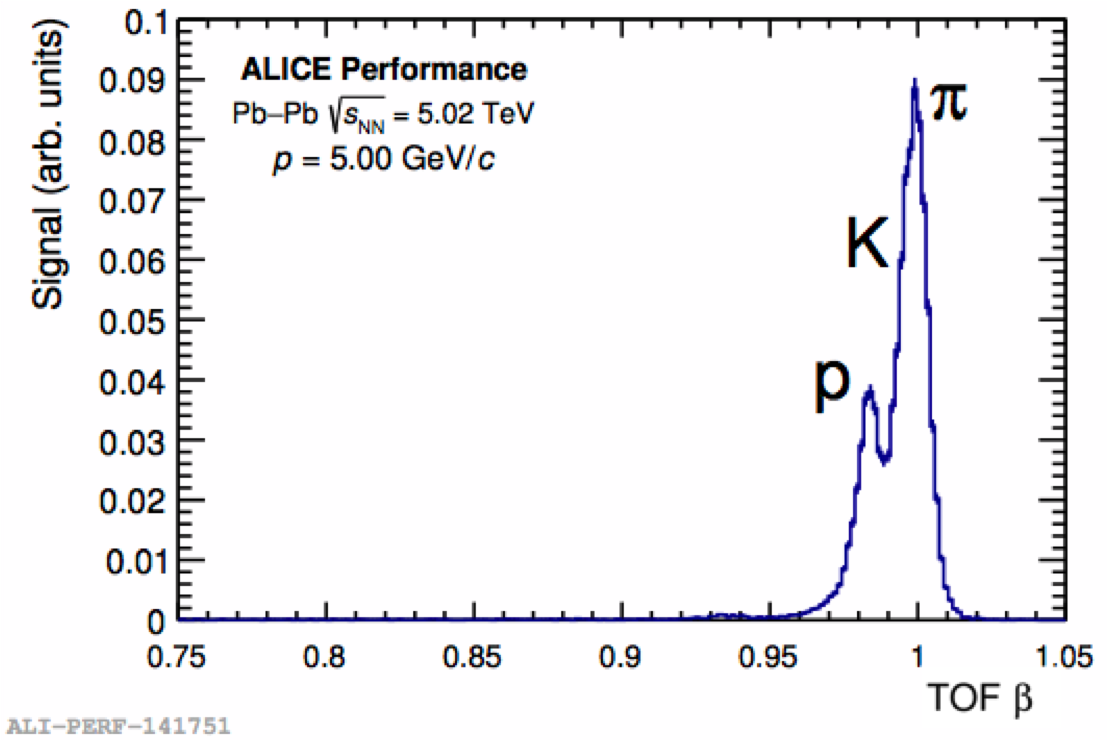}}}
        \caption{(a) Particle velocity ($\beta=$ v/$c$) measured by the TOF as a function of momentum in Pb--Pb collisions at $\sqrt{s_{\text{NN}}}=5.02~\text{TeV}$ after the calibration procedures. Points corresponding to non physical values are due to mismatched tracks at TOF, e.g. hit registered at TOF associated to the wrong track. (b), (c) Distributions of the $\beta$  for a fixed momentum of $3~\text{GeV}/c$ and $5~ \text{GeV}/c$, respectively.\label{fig:runpid}}
\end{figure}

Figure \ref{fig:TOFBetawoMismatchEtaCut_PbPb} shows the particle velocity ($\beta=$ v/$c$) measured by the TOF as a function of the momentum. The electron, pion, kaon, proton and deuteron bands are observed to be separated.
The PID estimator $\text{n}_{\sigma}$ can be expressed as
\begin{equation}
\label{eq:tTOFPID}
\text{n}_{\sigma}=\frac{t_{\text{TOF}}-t_{\text{event}}-t_{\text{exp}_\text{i}}}{\sigma_{\text{TOT}}}
\end{equation} 
As explained in \cite{TOFperform2013}, the better the time resolution the better the TOF PID capability: at a fixed $\text{n}_{\sigma}$, a good PID can be obtained up to a higher momentum. 
Thanks to the improved time resolution, the TOF can now provide a K/$\pi$ separation (see Fig.\ref{fig:TOFBeta_Cut3_PbPb}) up to 3 GeV/$c$ (instead of the design 2.5 GeV/$c$) and a p/K separation (see Fig.\ref{fig:TOFBeta_Cut5_PbPb}) up to  5 GeV/$c$ (instead of 4 GeV/$c$).

\section{Conclusions}
The ALICE TOF detector, after 10 years from the LHC start, shows very stable operations and no performance degradation neither at the expected Run 3 and Run 4 conditions. 
Thanks to a new campaign of calibration, the time resolution has been improved, reaching the record value of 56 ps. Consequently the TOF event time resolution improved, reaching the value of 10 ps with 100 tracks. The detector can now provide a K/$\pi$ separation up to 3 GeV/$c$ and a p/K separation up to  5 GeV/$c$. The TOF PID is therefore extensively and successfully exploited in many analyses in ALICE.
\bibliographystyle{utphys}
\bibliography{biblio}{}

\end{document}